# Neutron-hole states in $^{45}$Ar from p($^{46}$Ar,d)$^{45}$Ar reactions

F. Lu (卢飞)[1,2,+], Jenny Lee(李曉菁)[2,3], M.B. Tsang(曾敏兒)[2*], D. Bazin[2], D. Coupland[2], V. Henzl[2], D. Henzlova[2], M. Kilburn[2], W.G. Lynch(連致標)[2], A.M. Rogers[2], A. Sanetullaev[2], Z.Y. Sun (孙志宇)[2,4], M. Youngs[2], R.J. Charity[5], L.G. Sobotka[5], M. Famiano[6], S. Hudan[7], M. Horoi[8], Y. L. Ye (叶沿林)[1]

[1]*School of Physics and State Key Laboratory of Nuclear Physics and Technology, Peking University, Beijing 100871, China*

[2]*National Superconducting Cyclotron Laboratory, Michigan State University, E. Lansing, MI 48824, USA*

[3]*RIKEN Nishina Center, Hirosawa 2-1, Wako Saitama 351-0198, Japan*

[4]*Institute of Modern Physics, CAS, Lanzhou 730000, China*

[5]*Department of Chemistry, Washington University, St. Louis, MO 63130, USA*

[6]*Department of Physics, Western Michigan University, Kalamazoo, MI 49008, USA*

[7]*Department of Chemistry, Indiana University, Bloomington, IN 47405, USA*

[8]*Department of Physics, Central Michigan University, Mount Pleasant, MI 48859, USA*

## Abstract

To improve the effective interactions in the pf shell, it is important to measure the single particle- and hole- states near the N=28 shell gap. In this paper, the neutron spectroscopic factors of hole-states from the unstable neutron-rich $^{45}$Ar (Z=18, N=27) nucleus have been studied using $^{1}$H($^{46}$Ar, $^{2}$H)$^{45}$Ar transfer reaction in inverse kinematics. Comparison of our results with the particle-states of $^{45}$Ar produced in $^{2}H(^{44}Ar, H)^{45}$Ar reaction shows that the two reactions populate states with different angular momentum. Using the angular distributions, we are able to confirm the spin assignments of four low-lying states of $^{45}$Ar. These are the ground state ($f_{7/2}$), the first-excited ($p_{3/2}$), the $s_{1/2}$ and the $d_{3/2}$ states. While large basis shell model predictions describe spectroscopic properties of the ground and $p_{3/2}$ states very well, they fail to describe the $s_{1/2}$ and $d_{3/2}$ hole-states.

**\*** Corresponding author: tsang@nscl.msu.edu
+Present address: Shanghai Institute of Applied Physics, CAS, Shanghai 201800, China

# I. Introduction

The properties of low-lying single-particle states in nuclei near closed shells are key to the reduction of complex many-body physics of nuclei to the shell model consisting of valence nucleons and an inert core. To describe the observed nuclear structure, it is common to tune the nuclear Hamiltonian to the known properties of single-particle or single-hole states in the region. Currently, there is limited knowledge for many of the thousands of nuclei away from stability, particularly on the neutron-rich side where large changes in the shell model are expected [1,2]. Direct reactions provide a tool to understand details of nuclear shell structure as they probe single particle- or hole-states [3,4].

Unlike the sd shell [5], residual interactions in the pf shell are less comprehensive [6,7]. Due to mixing of the $p_{3/2}$ orbit outside the $f_{7/2}$ orbit in the N=28 core, shell model predictions of the excited states in pf-shell nuclei are less reliable. For example, predictions of the neutron spectroscopic factors (SF) for the excited states of Ni isotopes are not better than a factor of two[8,9]. Our previous systematic studies of the neutron spectroscopic factors of the excited states mainly use particle-states [8,9] from (d,p) transfer reactions. In order to obtain a complete understanding how single-particle states evolve from stability to instability, it is important to measure hole-states as well as particle-states [10]. For clarity, we use the convention of ref. [11] that the spectroscopic factors from stripping reactions, such as (d, p) reaction, which populate particle-states, are referred to as $S^+$ while spectroscopic factors from pickup reaction, such as (p, d) reactions, which preferentially populate hole-states, are referred to as $S^-$.

The particle-states of $^{45}$Ar have been studied using the $^2$H($^{44}$Ar, H)$^{45}$Ar transfer reaction in ref. [12]. In this paper, we report our findings on the neutron spectroscopic factors ($S^-$) of four states of $^{45}$Ar populated using the pick-up reaction, H($^{46}$Ar, $^2$H)$^{45}$Ar. To provide a simple orbital configuration of these states, we assume that $^{45}$Ar is produced from a one-step direct reaction mechanism that the (p,d) reaction removes one neutron from $^{46}$Ar. The ground state of $^{45}$Ar corresponds to a neutron-hole in the $f_{7/2}$ orbit. Two states with excitation energy around 1.75 MeV appear as a doublet in the deuteron energy spectra. Each of these states corresponds to a neutron-hole in the $s_{1/2}$ orbit or the $d_{3/2}$ orbit below the Fermi surface. However, the mechanism in populating the first-excited state at 0.542 MeV with a $p_{3/2}$ orbit is more complex. A one-step direct reaction requires removal of a $p_{3/2}$ neutron from a configuration of $^{46}$Ar with 2 neutrons in the $p_{3/2}$ orbit. As this state is also populated by addition of a neutron to the $p_{3/2}$ orbit via the $^2H(^{44}Ar, H)^{45}$Ar reaction, the wavefunction of this state may contain contributions from both particle- or hole- configurations. An alternative two-steps mechanism to produce the first excited state of $^{45}$Ar involves removal of a neutron in the $f_{7/2}$ orbit followed by promotion of the unpaired neutron in $f_{7/2}$ to $p_{3/2}$ orbit.

## II. Experimental Setup

The experiment was performed at the National Superconducting Cyclotron Laboratory at Michigan State University [13,14]. Radioactive secondary beam of $^{46}$Ar at 33 MeV/nucleon was produced in the Coupled Cyclotron Facility, and focused on a polyethylene target, $(CH_2)_n$, with areal densities of 2.29 mg/cm$^2$. To allow clean identification of the deuterons from the desired reactions, both the deuteron and $^{45}$Ar were detected in coincidence, allowing a kinematically complete measurement. The deuterons were detected in the High-Resolution Array (HiRA) [15] and the recoil residues were detected in the focal plane of the S800 mass spectrometer [16].

An array of 16 HiRA telescopes was placed at 350 mm from the target where they subtended polar angles of 4°- 45° in laboratory frame. Each telescope contained a 65 μm thick ΔE and a 1500 μm thick E silicon strip detectors, backed by four 39 mm thick CsI(Tl) crystals. The 32x32 strips in the double-sided E detector effectively subdivided each telescope into 1024 2mm x 2mm pixels, each with an angular resolution of ±0.16°. The HiRA array was placed in a vacuum chamber in front of the S800 mass spectrometer.

Due to forward focusing of the emitted deuterons in an inverse kinematic measurement, our setup covered most of the relevant solid angles. The average geometrical efficiency was about 30% over the covered angular domain. To ensure excellent energy and angular resolutions, the position of every pixel relative to the target was determined to sub-millimeter accuracy using the Laser Based Alignment System (LBAS) [17]. In this experiment, an accuracy of approximate 0.3 mm, which corresponded to $0.05°$ in the alignment of the detectors and the reaction target, was achieved [17].

Deuterons were identified in HiRA with standard energy-loss techniques using the energy deposited in the ΔE and E Silicon-strip and CsI detectors. Reaction residues were identified in the S800 mass spectrometer using the energy loss in the ionization chamber and the time-of-flight between the cyclotron radio frequency and the focal plane detector. The S800 magnetic fields were set to optimize selection of the $^{45}$Ar residues. Detailed descriptions of the experimental setup and analysis can be found in [14,18].

## III. Excitation-energy spectra of $^{45}$Ar

Figure 1 shows the reconstructed excitation-energy spectra of $^{45}$Ar from deuterons emitted at laboratory angles forward of 19° where the deuteron energy is low enough to stop in the Si E detectors. Measurements using a 17 mg/cm$^2$ carbon target reveal that the background from the C nuclei in the polyethylene target is negligible when both the deuteron and the heavy recoil nucleus are detected in coincidence. Well-defined peaks up to 4.5 MeV, as shown in Fig. 1, are observed. We are able to identify seven peaks in the energy spectrum. Two of the peaks, at 1.735 MeV and 1.770 MeV form an unresolved doublet, in agreement with previously measured states compiled by the National Nuclear Data Center, NNDC [19, 20]. These peaks are also listed in Table 1. The dashed curves are individual Gaussian fits to the peaks and the solid curve is the sum of all the Gaussian fits. The excess counts around 1.3 MeV is suggestive of a peak. The quality and the statistics of our data, however, does not allow for a definitive determination. From the fits, we obtain the FWHM of the ground-state peak to be 450 keV and 380 keV for the strong peak around 1.75 MeV. The worsening of the energy resolution of the ground state peak is due to kinematic broadening at larger angles. As explained below, the main contribution to the 1.74 MeV peak comes from $2s_{1/2}$ state which is forward peak at zero degree with little kinematic broadening while the $f_{7/2}$ ground state contribution has a rather flat contribution below 20°. GEANT4 simulations suggest that the largest contributions to the energy resolution come from the relatively thick target (2.29 mg/cm$^2$) used in the experiment [18].

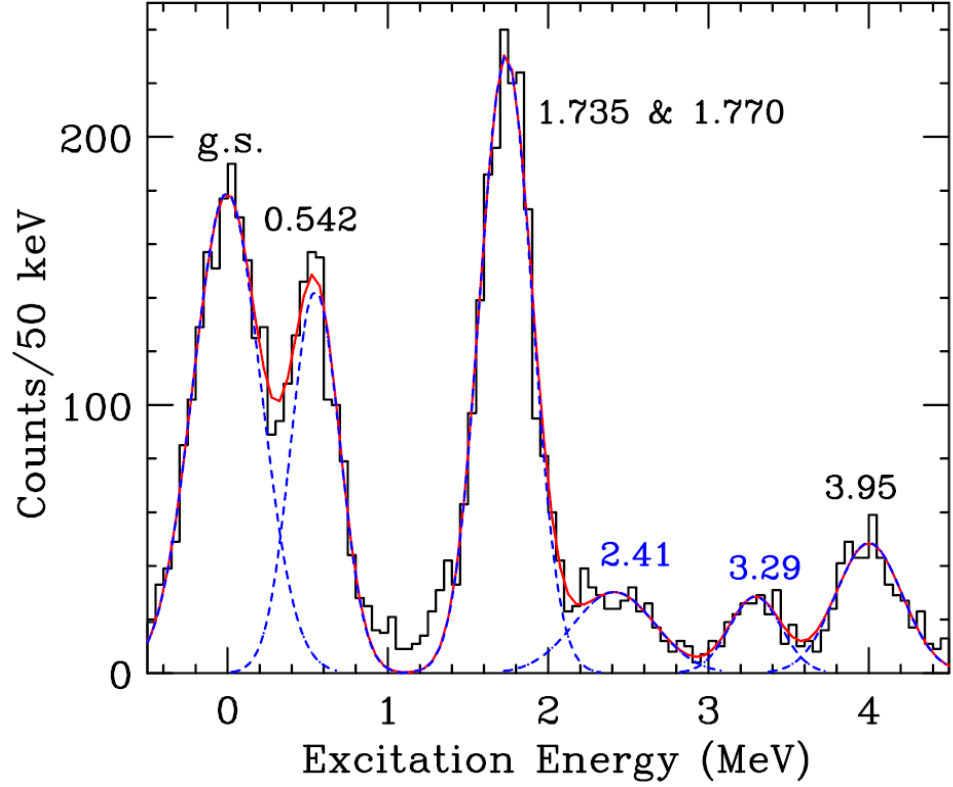

**Figure 1:** (Color online) Excitation-energy spectra of $^{45}$Ar reconstructed from the energy and angles of the deuterons detected in coincidence with $^{45}$Ar in the H($^{46}$Ar, $^{45}$Ar)$^2$H reaction. To optimize the energy resolution, only data with $\theta_{c.m.} = 4°-19°$ where deuterons that stop in the Si E detectors are included. The excited state energies are labeled inside the figure and given in units of MeV.

## IV. Angular Distributions

To identify the spin and parity of the observed states, angular distributions which give information on the angular momentum of the state have been measured. In figure 2, we show the angular distributions of the differential cross-sections from the strongest four peaks (E*~0, 0.542, 1.75 and 3.95 MeV) identified in Figure 1. The ground state and first-excited state (0.542 MeV) are not completely resolved. However, at forward angles where kinematic broadening has the least effects, we can distinguish the peaks of the ground and first-excited states at small angles up to 13° and extract differential cross sections using double-Gaussian fits. These cross sections are presented in Figures 2a and 2b. Different *l* values for the ground state (*l*=3) and first-excited state (*l*=1) result in very different angular distributions. Figure 2(b) shows that the calculated angular distributions for the excited state with *l*=1 drops sharply at $\theta_{c.m.}$>10°. Contributions from the first-excited states to the data points at $\theta_{c.m.}$ = 20.9°, 23.3° and 26.3° are much less than 10%. The data plotted at the large angles in Figure 2(a) reflect mainly the cross-sections from the ground state.

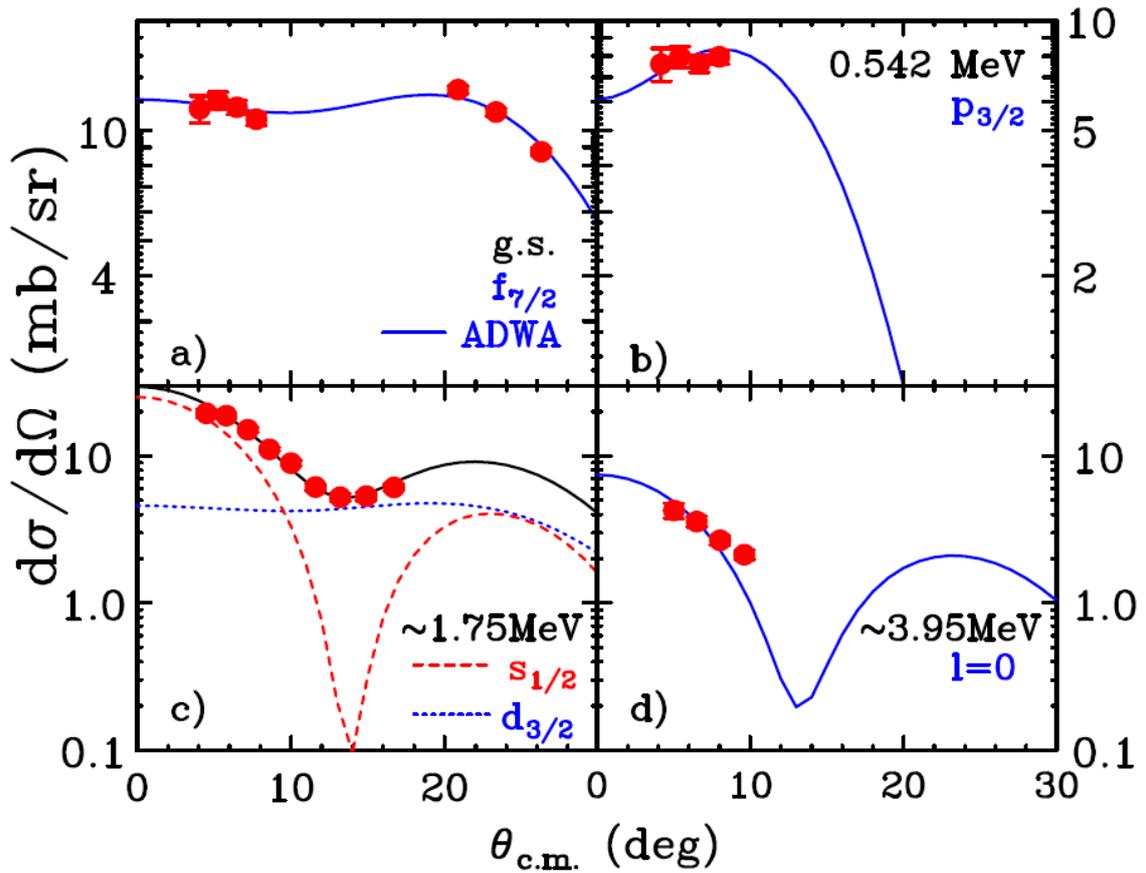

Figure 2: (Color online) Differential cross-sections from H($^{46}$Ar, $^{2}$H)$^{45}$Ar reaction in the center-of-mass frame for transitions to (a) ground state, (b) first-excited state at 0.542 MeV, (c) doublet states around 1.75 MeV, and, (d) a state at 3.95 MeV. The red solid circles are the data and the curves are calculations from ADWA normalized by the spectroscopic factors. In (c), the dashed and dotted curves correspond to $2s_{1/2}$ and $1d_{3/2}$ transitions, respectively, and the solid curve is the sum of the two calculations.

To extract the spectroscopic factors, we fit the angular distributions in Figure 2(a) using the Adiabatic Distorted Wave Approximations (ADWA) as described in ref. [21, 22, 23]. In the case of (p,d) and (d,p) transfer reactions where the deuteron is weakly bound and breaks up easily, scattering in the reaction, $A+a \rightarrow B+b$, (where a=p and b=d) in (p,d) reactions and (a=d, b=p) in (d,p) reactions), can be described in the three-body ADWA model. The adiabatic approximation is incorporated in the distorted-wave Born approximation (DWBA) framework to take into account the influence of deuteron break-up. The reaction is regarded as a perturbation to the elastic scattering that induces a transition to occur between two channels. The relative motions are governed by an optical potential. The resulting waves, distorted from the inelastic-scattering plane waves, are used to obtain the approximate transition amplitude which is a nuclear overlap function and subsequently calculate the differential cross sections. In the systematic studies of the neutron spectroscopic factors for nuclei with Z<28 [8, 22, 23] using nucleus-nucleus global optical potentials, Chapel-Hill89 [24], is found to provide consistent neutron spectroscopic factor values. For the bound neutron, fixed values of the radius parameter of 1.25 fm and diffuseness parameter of 0.65 fm are used. The depths of the central potential wells are adjusted to reproduce the experimental binding energies. The TWOFNR code from the University of Survey is used to provide the reaction calculations [25]. One advantage of this algorithm is the good agreement between the experimental and shell model spectroscopic factors which provide insights into the nuclear structure and limitation of modern day shell models. However, shell models do not include all the nucleon correlations which tend to quench the SF values. Since the extracted SF values depend on input parameters to the reaction model [23, 26], the SF values presented in the present work can be used to compare to SF values of neighboring nuclei obtained with the same input parameters but may not be considered as the absolute SF values.

The fitted angular distributions are shown as solid lines in Figure 2. The normalization factors that fit the calculated angular distributions to the data are the spectroscopic factors. The deduced spectroscopic factors ($S^-$) is 5.29$\pm$0.40 for the ground state which is consistent with the value of 5.08$\pm$0.30 obtained in ref. [14] where the ground-state spectroscopic factor is extracted by fitting the angular distributions of the combined peaks corresponding to states at 0 and 0.542 MeV. Shell model predictions for the ground state $S^-$ value is 5.34 using the SDPF-U interaction [27] in the sd-pf model space (with the code Antoine[28] or NuShellx [29]). The extracted spectroscopic factors are in reasonable agreement with shell model predictions as discussed in ref [8,9,22,23]. For the ground state, theoretical uncertainties from the reaction model up to 11% using the Faddeev calculations have been estimated in ref. [30]. Determinations of similar theoretical uncertainties for excited states are not available. Since the theoretical uncertainties are very sensitive to the actual reactions and beam energy, it is difficult to estimate or extrapolate theoretical uncertainties using Faddeev calculations [30]. Thus the theoretical uncertainties are not included in the present work.

The deduced spectroscopic factors ($S^-$) are 0.51$\pm$0.06 for the first-excited state ($p_{3/2}$) at 0.540 MeV as compared to 0.59 from shell model predictions. As noted in ref. [31], shell model calculations reveal that this state has significant contributions from the excited core. More interestingly, the spectroscopic factor of the same $p_{3/2}$ state in $^{47}$Ca (which is two protons more than $^{45}$Ar), is a factor of 10 smaller [32-34]. This fact is reproduced by shell model as discussed in ref. [10].

A strong peak around 1.75 MeV is clearly seen in Figure 1. This prominent peak results from the contributions of the transitions leading to two different states with excitation energies of 1.735 MeV and 1.770 MeV [19,20] which cannot be resolved in the present work. The existence of a doublet state is consistent with the systematics of the N=27 isotones studied using the (p,d) pickup reactions. In the case of $^{47}$Ca, which has two more protons than $^{45}$Ar, a similar peak consisting of the doublet of $s_{1/2}$ (2.6 MeV) and $d_{3/2}$ (2.578 MeV) states is observed in $^{48}$Ca(p,d)$^{47}$Ca reaction [32-34].

The angular distribution for the doublet peak at ~1.75 MeV is shown in Figure 2(c). It is found that the angular-momentum transfers of $l$=0, 2 ($s$, $d$) orbits are needed in order to reproduce the angular distribution. The dashed curve corresponds to the calculated angular distributions of the $2s_{1/2}$ state which peak at zero degree. The dotted curve with relatively flat distributions over the angular range from 0° to 30°, corresponds to the calculated angular distributions of the $1d_{3/2}$ state. Using $\chi^2$ analysis by fitting the data, the extracted S$^-$ values are 0.77±0.08 and 2.32±0.23 for the $s_{1/2}$ and $d_{3/2}$ states respectively. (If one assumes no contributions from $d_{3/2}$ state, a maximum S$^-$ value of 1.2 is obtained but the fits are not very good.) The solid curve is the total contributions with each distribution multiplied by the corresponding spectroscopic factor values of 0.77 and 2.32. Predictions for hole-states in the $1s_{1/2}$ and $0d_{3/2}$ orbits from shell model are not available at present due to lack of better effective interactions that can predict these states. The SDPF-U interactions [27] used to calculate the ground and first excited states was designed for $0\hbar\omega$ calculations where the sd orbits below the $f_{7/2}$ orbit are filled. It was recently extended to describe up to 6-particle 6-hole excitations from the sd orbits [35] but this version [36] is not yet ready to describe the $1s_{1/2}$ and $0d_{1/2}$ hole-states. In the limit of independent-particle model, the spectroscopic factors will acquire the values of 2j+1 or 2 and 4 for the $s_{1/2}$ and $d_{3/2}$ states respectively. Thus the lowest identified s and d states in the present work contribute nearly 50% of the spectroscopic strength.

We also analyze another group of states found at the excitation energy of 3.95 MeV, where there is no information listed in NNDC [19, 20] about the orbital configuration. The corresponding forward-angle differential cross sections can be clearly identified and are plotted in Figure 2(d). The shape of the present angular distributions are consistent with the $l=0$ calculations using ADWA, suggesting the existence of $l=0$ strength for the state at 3.95 MeV. The angular distributions, however, cannot be used to assign the spin and parity to this state. With a tentative $l=0$ assignment, the spectroscopic factor would be 0.15±0.03.

## V. Comparison of particle- and hole- states of $^{45}$Ar

Table I compiles the information about the states of $^{45}$Ar identified in the NNDC [19, 20], hole-states from the $^{1}H(^{46}Ar,^{2}H)^{45}Ar$ reaction obtained in the present work and particle-states from $^{2}H(^{44}Ar,H)^{45}Ar$ reaction published in ref. [12]. Since the energy levels in NNDC were obtained through gamma-ray studies, they are much more accurate than the energy levels obtained in transfer reactions, especially in inverse kinematics. We defer to the energy level given in NNDC. In both transfer reaction studies, the $l$ value of the ground state is unmistakably 3, corresponding to the $f_{7/2}$ valence-neutron configuration. Unlike the corresponding ground-state spectroscopic factor of $^{47}$Ca from $^{48}$Ca(p,d)$^{47}$Ca reactions, the obtained value of S$^{-}$(g.s.)=5.29±0.30 is below the independent-particle value of 8. The smaller S$^{-}$ value reflects evolution of the orbit under the influence of non-closed proton shell in Ar isotopes.

The first excited state of $^{45}$Ar, with a neutron in the $p_{3/2}$ orbit and two holes in the f$_{7/2}$ orbit, are populated in both the (p,d) and (d,p) reactions. In $^2H(^{44}Ar,H)^{45}Ar$ reaction, one can consider the neutron in the p$_{3/2}$ orbit as created by neutron addition and S$^+$(0.542 MeV)=0.76±0.20. Assuming a one-step neutron removal mechanism in $^1H(^{46}Ar, ^2H)^{45}Ar$ reaction, S$^-$(0.542 MeV)=0.51±0.06) is obtained for the neutron hole-state.

Figure 1 shows a hint of a state at ~ 1.34 MeV. We make a "?" mark next to this level as identification of this state is far from certain. Two groups of states around 2.41 MeV and 3.29 MeV can be seen clearly in our data. However the resolution at backward angles of these states does not allow us to extract their angular momemtum values. Thus we put "?" mark in the spin and parity assignment for these states.

Of 19 excited states of $^{45}$Ar below 4.9 MeV listed in Table I, seven states are populated in the $^1H(^{46}Ar,^2H)^{45}Ar$ while nine states are populated in the $^2H(^{44}Ar,H)^{45}Ar$ reactions. Four states (E*=1.660, 1.911, 2.757, 4.326 MeV) are not populated by neither transfer reactions. Except for the ground and first excited states, each reaction populates different states. Clearly, complete information of effective single-particle energy [10] requires the use of both pickup (p,d) and stripping (d,p) reactions to study the hole- and particle-states in $^{45}$Ar.

| NNDC | | $^1H(^{46}Ar, ^2H)^{45}Ar$ | | | | $^2H(^{44}Ar, H)^{45}Ar$ | | | |
|---|---|---|---|---|---|---|---|---|---|
| Level (keV) | $J^\pi$ | Level (keV) | l | $J^\pi$ | $S^-(^{46}Ar \rightarrow ^{45}Ar)$ | Level (keV) | l | $J^\pi$ | $S^+(^{44}Ar \rightarrow ^{45}Ar)$ |
| 0 | 5/2, 7/2- | 0 | 3 | 7/2- | 5.29±0.30 | 0 | 3 | 7/2- | 1.52±0.40 |
| 542.1 | 1/2-,3/2- | 540 | 1 | 3/2- | 0.51±0.06 | 550 | 1 | 3/2- | 0.76±0.20 |
| 1339.9 | | 1.34(?) | | | | | | | |
| 1416.1 | 1/2-,3/2- | | | | | 1420 | 1 | | 1.08±0.16 |
| 1660? | | | | | | | | | |
| 1734.7 | | 1750 | 2 | 3/2+ | 2.32±0.23 | | | | |
| 1770.3 | | 1750 | 0 | 1/2+ | 0.77±0.08 | | | | |
| 1876 | 1/2-,3/2- | | | | | 1880 | 1 | | 0.32±0.02 |
| 1911 | | | | | | | | | |
| 2420 | | 2.4 | | (?) | | | | | |
| 2510 | 1/2-,3/2- | 2.52 | | | | 2510 | 1 | | 0.46±0.06 |
| 2757 | | | | | | | | | |
| 3230 | | 3.27 | | | | 3230 | 1 | | |
| 3294.8 | | 3.29 | | (?) | | | | | |
| 3718 | | | | | | 3720 | 1 | | |
| 3949.7 | | 3.95 | 0 | | 0.15±0.03 | | | | |
| 4280 | | | | | | 4280 | 1 | | |
| 4326.1 | | | | | | | | | |
| 4880 | | | | | | 4770 | 3 | | 1.08±0.24 |

**Table I:** Energy levels (<5 MeV), spin and parity assignments compiled by the National Nuclear Data Center [19, 20]. The hole-states and the corresponding spectroscopic factors, S⁻ determined from removal of a valence neutron, in this work are listed in the three columns under $^1H(^{46}Ar, ^2H)^{45}Ar$ reaction. The particle levels and the corresponding S⁺ values determined from addition of a valence neutron [12] are listed in the right most three columns under the $^2H(^{44}Ar, H)^{45}Ar$ reaction.

## VI. Summary

In summary, the (p,d) neutron pick up and (d,p) neutron striping reactions populate different states. Complete information of the hole- and particle- states in $^{45}Ar$ requires the use of both reactions in inverse kinematics.

The neutron spectroscopic factors for the neutron hole-states are 5.29±0.30 ($f_{7/2}$ ground state), 0.51±0.06 ($p_{1/2}$, first-excited state), 0.77±0.08 ($s_{1/2}$) and 2.32±0.23 ($d_{3/2}$). The deviation of the ground state spectroscopic factor from the independent particle of 8 reflects evolution of the $f_{7/2}$ state from $^{47}$Ca with Z=20, a magic spherical core) to a deformed core in $^{45}$Ar with Z=18. The excellent agreement with shell model predictions of the ground and first excited states indicate success of the modern interactions used to describe the $f_{7/2}$ and $p_{3/2}$ orbits in the pf shell. However, neither the energy levels nor the spectroscopic factors of the $s_{1/2}$ and $d_{3/2}$ deep-hole states observed in $^{47}$Ar can be explained by shell model calculations suggesting that the work to improve the effective interaction to describe transitions from sd to pf orbits is far from complete.

**Acknowledgement**

This work is partly supported by the National Science Foundation Grants No. PHY-0606007, PHY-1068217 (MH) and by the National Science Foundation of China Grants No. 11035001 FL acknowledges support of Michigan State University during his stay at the NSCL.